\documentclass[runningheads]{llncs}

\usepackage{tabularx}
\usepackage{amsmath}
\usepackage{graphicx,subfigure}
\usepackage{marvosym}

\usepackage{xcolor}
\newcommand{\shuiqiao}[1]{\textcolor{black}{#1}}
\newcommand{\dahuihui}[1]{\textcolor{black}{#1}}

\begin{document}
%
\title{Detecting Topic and Sentiment Dynamics Due to COVID-19 Pandemic Using Social Media}
\titlerunning{Detecting Topic and Sentiment Dynamics Due to COVID-19 ...}
%
\author{Hui Yin\inst{1} \and
Shuiqiao Yang\inst{2} \and
Jianxin Li\inst{1}\textsuperscript{\Letter}}
\authorrunning{Hui Yin et al.}
%
\institute{School of Information Technology, Deakin University, Geelong, Australia 
\and Data Science Institute, University of Technology Sydney, Sydeny, Australia \\
\email{jianxin.li@deakin.edu.au}}
\maketitle              
\begin{abstract}
The outbreak of the novel Coronavirus Disease (COVID-19) has greatly influenced people's daily lives across the globe. Emergent measures and policies (e.g., lockdown, social distancing) have been taken by governments to combat this highly infectious disease. However, people's mental health is also at risk due to the long-time strict social isolation rules. Hence, monitoring people's mental health  across various events and topics will be extremely necessary for policy makers to make the appropriate decisions.
On the other hand, social media have been widely used as an outlet for people to publish and share their personal opinions and feelings. 
The large scale social media posts (e.g., tweets) provide an ideal data source to infer the mental health  for people during this pandemic period. In this work, we propose a novel framework to analyze the topic and sentiment dynamics due to COVID-19 from the massive social media posts. 
Based on a collection of 13 million tweets related to COVID-19 over two weeks, we found that the positive sentiment shows higher ratio than the negative sentiment during the study period.
When zooming into the topic-level analysis, we find that different aspects of COVID-19 have been constantly discussed and show comparable sentiment polarities. 
Some topics like ``stay safe home" are dominated with positive sentiment. The others such as ``people death" are consistently showing negative sentiment. Overall, the proposed framework shows insightful findings based on the analysis of the topic-level sentiment dynamics.



\end{abstract}

\section{Introduction}
\label{Introduction}

The outbreak of the novel Coronavirus Disease 2019 (COVID-19) has influenced millions of people around the world \cite{noauthor_coronavirus_2020}. People's lives are at risk due to the fact that COVID-19 is highly infectious from person to person. It is reported that the globally confirmed cases of COVID-19 have surpassed 10 milloin and more than 500,000 people have lost their lives due to the infection until 29 June 2020 \footnote{\url{https://news.google.com/covid19/map?hl=en-AU\&gl=AU\&ceid=AU\%3Aen}}.
Many countries and jurisdictions have taken a series of specific measures to help to slow down the spread of COVID-19, such as travel ban, lockdown, closure of public places (e.g., gym, restaurants, schools), requiring people to practise good personal hygiene, keep physical social distancing of 1.5 meters, and work or study at home to reduce contact with others.
The above measures have changed people's daily lives, and most people have to follow the policies to protect the safety of their communities.
However, many mental symptoms like worry, fear, frustration, depression and anxiety could occur and cause serious mental heath issues to people due to the long-time social activity restriction during the pandemic period \cite{zhou2020covid19}. 
Therefore,  understanding when and where people would experience mental issues in this special period is important for governments to make the right decisions. 

On the other hand, people could spend more time on social media platforms like Twitter to get the latest news, communicate with friends and \dahuihui{post} their feelings and thoughts during the lockdown period.
Such massive personal posts from social media could become invaluable data sources for large-scale sentiment and topic mining for monitoring people's mental health  across different events or topics \cite{Yang2019}. 
Hence, many recent work focuses to detect the social sentiment for people due to COVID-19 \cite{zhou2020covid19,han2020using,prabhakar2020informational}. 
For instance, Zhou et al. \cite{zhou2020covid19} adopted Twitter data for massive sentiment analysis due to COVID-19 for people living New South Wales Australia. Han et al. \cite{han2020using} studied the public opinion in the early stages of COVID-19 in China by analyzing text data from Sina Weibo. Rajesh et al. \cite{prabhakar2020informational} exploited topic model to generate topics from tweets related to coronavirus and calculated the presence of different emotions. 
However, little work is found to analyze the topic and sentiment dynamics together, even though such dynamical analysis is important for authorities to understand when and where people would experience mental health issues.

In this work, we aim to dynamically identify the popular topics and their associated sentiment polarities due to the COVID-19 pandemic.
Our research questions are: (1) How is the dynamics of people's sentiment? (2) Which topics are mostly discussed by people? (3) How is the evolution of different topics? (4) How is the sentiment dynamics of topics? 
To answer these questions, we propose a novel dynamic topic discovery and sentiment analysis framework which contains multiple modules include data crawling, data cleaning, topic discovery, sentiment analysis and result visualization. 
In the proposed framework, we employ the Dynamic Topic Model (DTM) \cite{blei2006dynamic} to generate accurate daily topics. To determine the sentiment polarity of each topic and tweet, we utilize a sentiment lexicon tool: VADER\cite{hutto2014vader} \footnote{https://github.com/cjhutto/vaderSentiment} to infer the sentiment polarity. 
We collect 13,746,822 tweets from 1 April 2020 to 14 April 2020 related to COVID-19 from Twitter across the world to test the effectiveness of the proposed framework.
The experimental results  show that the proposed framework can generate insightful findings such as the overall sentiment dynamics among people, topic evolutionary patterns, the sentiment dynamics of different topics. 
\section{Related work}
With the spreading of COVID-19 across the world, researchers have proposed to use sentiment analysis based on social media as a tool to monitor people's mental health. In this section, we review the latest work related to COVID-19 analysis \dahuihui{based on social media data}. 


Rajesh et al.\cite{prabhakar2020informational} adopted a classic Latent Dirichlet Allocation (LDA) topic model method to generate 10 topics in a random sample of 18,000 tweets about coronavirus, then they used NRC sentiment dictionary to calculate the presence of eight different emotions, which were ``anger", ``anticipation", ``disgust", ``fear", ``joy", ``sadness", ``surprise" and ``trust".  
Han et al. \cite{han2020using} explored public opinion in the early stages of COVID-19 in China by analyzing Sina-Weibo texts. 
The COVID-19 related micro-blogs were generalized into 7 topics and 13 more detailed sub-topics. However, they judged the polarity of the  topics according to the polarity of the topic words and failed to consider the specific tweets under this topic.
Cinelli et al. \cite{cinelli2020covid} analyzed engagement and interest in the COVID-19 topic on different social media platforms such as Twitter, Instagram, Reddit, and provided a diﬀerential assessment of the global discourse evolution of each platform and their users. They found that reliable and suspicious information sources have similar spreading patterns.
Depoux et al. \cite{depoux2020pandemic} confirmed that the spread of social media panic is faster than that of COVID-19. Therefore, the public rumors, opinions, attitudes and behaviors surrounding COVID-19 need to be quickly detected and responded to. They suggested to create an interactive platform dashboard to provide real-time alerts of rumors and concerns about the spread of coronavirus worldwide, which would enable public health officials and relevant stakeholders to respond quickly with a proactive and engaging narrative that can mitigate misinformation.
Sharma et al. \cite{sharma2020coronavirus} designed a dashboard to track misinformation in Twitter conversations. The dashboard allows visualization of the social media discussions around coronavirus and the quality of information shared on the platform and updated over time. They evaluated sentiment polarity for each tweet based on its textual information and showed the distribution of sentiment in different countries over time. 

More recently, Huang et al. \cite{huang2020disinformation} examined the public discussion concerning COVID-19 on Twitter and found that the most influential tweets are still written by regular users, such as news media, government officials, and individual news reporters. They also discovered that ``fake news" sites are most likely to be retweeted within the source country and so are less likely to spread internationally. 
Zhou et al. \cite{zhou2020covid19} exploited tweets on Twitter to analyse the sentiment dynamics of people living in the state of New South Wales (NSW) in Australia during the pandemic period. They first summarized that the overall polarity of the community since the outbreak was positive, and then analyzed the sentiment dynamics of the NSW local government areas in terms of lockdown, social distance and JobKeeper.

Different from the above work that either performed static sentimental analysis or failed to analysis the detailed topic-level sentiment. We propose to analyse the dynamics of topic-level sentiment to monitor the evolution of people's mental states across different topics or events.

\section{Proposed Framework}
\label{ProposedMethod}
\begin{figure}
    \centering
    \includegraphics[width=90mm, scale=0.9]{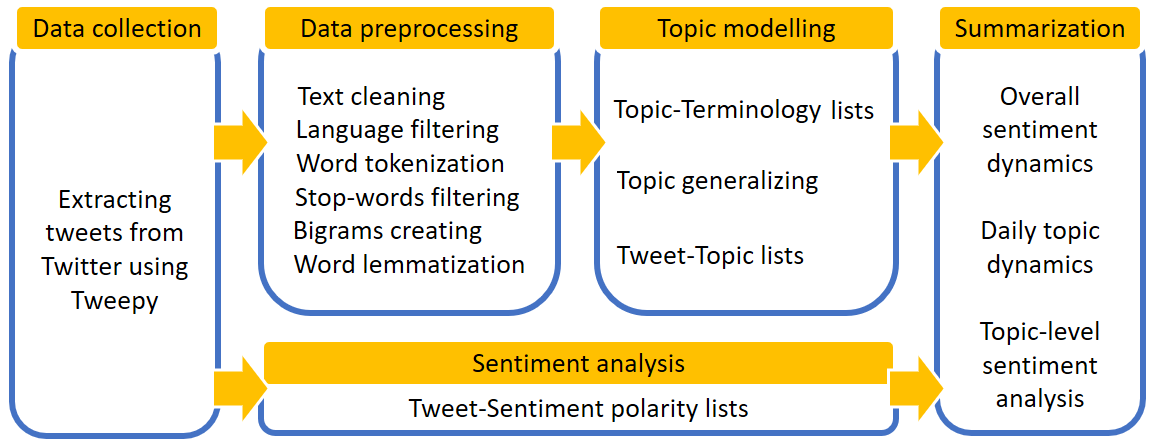}
    \caption{Overall procedure of analysis.}
    \label{Framework}
\end{figure}

In this section, we introduce the proposed framework and the adopted techniques for detecting topic and sentiment dynamics. Our framework  contains three steps for the task. Firstly, we divide the tweets into different topics generated by a dynamic topic model, then we determine the sentiment polarity of each tweet, and finally we summarize the sentiment polarity distribution of each topic. 
The proposed framework is shown in figure \ref{Framework}. 
\subsection{Topic Extraction}
Topic modelling is the process of learning, recognizing, and extracting high-level semantic topics across a corpus of unstructured text. A popular method for this is Latent Dirichlet Allocation (LDA)  proposed by Blei et al. \cite{blei2003latent}, which is used to detect topics for static corpus. 
Later, a Dynamic Topic Models (DTM) \cite{blei2006dynamic} is proposed based on LDA to  mine topic evolution over time by extending the idea of LDA to allow topic mining over fixed time intervals.
Hence, DTM can handle sequential documents and generate topics for different time slices of corpus.
Specifically, the documents within each time slice are modeled with the static LDA, the topics associated with slice $t$ evolve from the topics associated from the previous time slice $t-1$.
The dynamics of the topic model is given by Eq. \ref{DTMFormula} and 2, table \ref{notation} shows the meaning of the notations. 
DTM mines topics of each time slice with LDA and its parameters $\beta$ and $\alpha$ are chained together in a state space model which evolve with Gaussian noise to get a smooth evolution of topics from time to time.

\begin{table}[]
\centering
\caption{Meaning of the notations.}
\label{notation}
\begin{tabular}{l|l}
\hline
Symbol & Description                                                            \\ \hline
$\alpha_{t}$ & as the per-document topic distribution at time t.          \\ \hline
$\beta_{t,k}$ & as the word distribution of topic k at time t.             \\ \hline
$\eta_{t,d}$ & as the topic distribution for document d in time t.        \\ \hline
$z_{t,d,n}$ & as the topic for the $n^{th}$ word in document d in time t. \\ \hline
$\omega_{t,d,n}$ & as the $n^{th}$ word at time slice $t$, document d.                                      \\ \hline
\end{tabular}
\end{table}

\begin{align}
\label{DTMFormula}
    \beta_{t,k}\mid\beta_{t-1,k}\sim N(\beta_{t-1,k},\sigma_2 I )\\
    \alpha_{t}\mid\alpha_{t-1}\sim N(\alpha_{t-1},\delta_2I)
\end{align}



\subsection{\dahuihui{Tweets Sentiment Analysis}}
\begin{figure}[ht]
    \centering
    \includegraphics[width=115mm, scale=1.15]{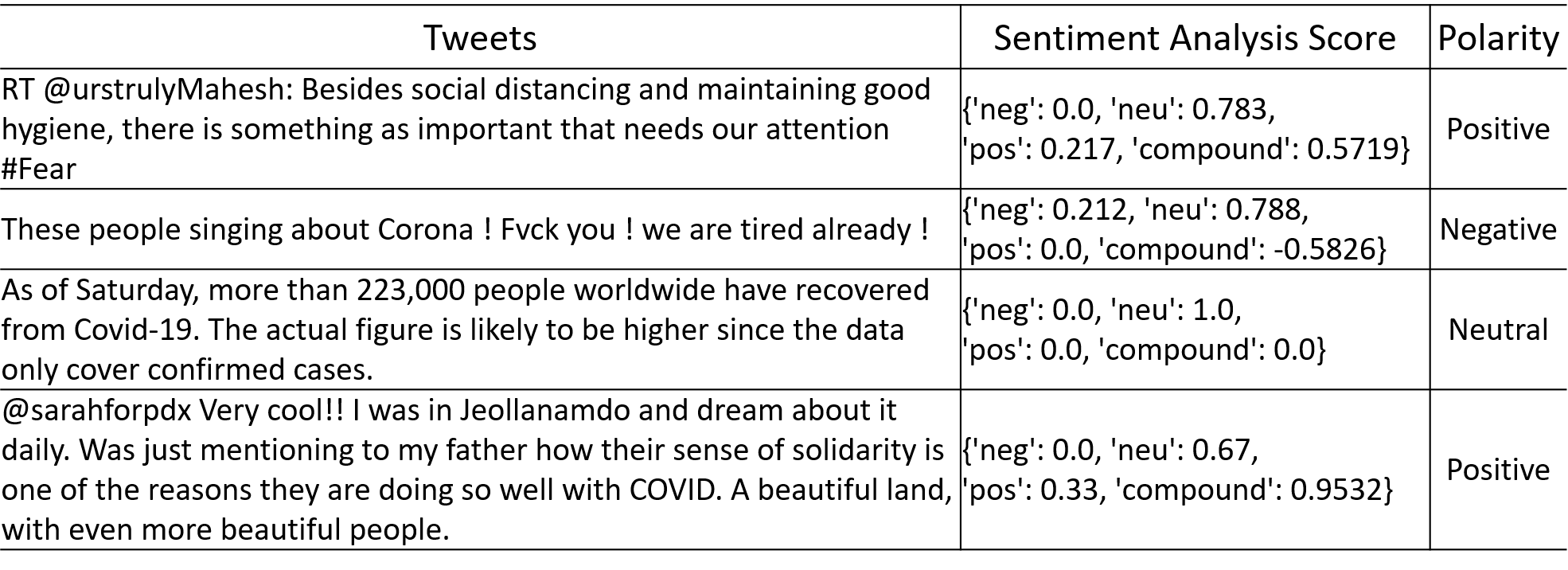}
    \caption{\dahuihui{Example tweets with sentiment polarity inferred by VADER.}}
    \label{TweetExample}
\end{figure}

Sentiment Analysis (SA) also commonly referred as Opinion Mining(OM) that aims to find opinionated information and detect the sentiment polarity. 
Nowadays, SA techniques are quite mature and many tools are openly available, such as Stanford's CoreNLP \cite{manning2014stanford}, VADER \cite{hutto2014vader}, SentiStrength \cite{thelwall2010sentiment}, SentiCircles \cite{saif2014senticircles}, which are specifically designed for social media conversation.
In this work, we employ VADER  to identify the sentiment polarity of each tweet. 
VADER (Valence Aware Dictionary and sEntiment Reasoner) is a lexicon and rule-based sentiment analysis tool that is specifically \shuiqiao{attuned} to sentiment expressed in social media. 
It was introduced by Hutto et al.in 2014 and has been widely used in many social media text sentiment analysis tasks  \cite{davidson2017automated,cheng2017anyone,ferrara2015measuring,you2016cross,sharma2020coronavirus,zhou2020covid19}.
VADER can classify the sentiment into negative, neutral, and positive categories and employ compound score which is computed by summing the valence scores of each word in the lexicon and normalized in range (-1,1), where ``-1" represents most extreme negative and ``1" represents most extreme positive.
If compound score is greater than 0.05, the text is considered positive, if the score is less than -0.05, it is considered negative, if the score is between 0.05 and -0.05, the text polarity is neutral.
The biggest advantage of VADER is that it does not require data preprocessing and model training, and can  be used directly on the raw tweet to generate sentiment polarity. Some examples of tweet sentiment results from VADER are shown in figure \ref{TweetExample}. 
In this work, we use VADER to produce the sentiment polarity of each original tweet, namely positive, negative and neutral. And then, combine the topic mining result from DTM to analyze the topic-level sentiment. 

\subsection{Measuring Topic Sentiment}
After the above two steps, all tweets will be clustered in the corresponding topics, and marked with sentiment polarity.
The sentiment polarity is aggregated over tweets to estimate the overall sentiment distribution.
For each topic per day, we sum up the number of positive, negative and neutral tweets in the topic, thus, each topic is associated with three sentiment counts. 

\begin{align}
     \label{sentiment}
      DT_{i,j} = N_p + N_n + N_o \\
      \label{totaltweet}
    N=\sum_{i=1}^{|D|}\sum_{j=1}^{|T|}DT_{i,j},
\end{align}

For each topic in the studied days, we define a sentiment distribution in Eq. \ref{sentiment} and 4. $|D|$ is the total studied days and $|T|$ is the total topic number. $DT_{i,j}$ represents the total number of tweets under topic $j$ in day $i$, $N_p$, $N_n$ and $N_o$ respectively denote the positive, negative and neutral tweet counts. $N$ represents the total number of tweets in our dataset. 
We observe the daily hot topics about COVID-19 on Twitter (April 1 to April 14), and analyze the sentiment polarity distribution of each topic.
The details are presented in Section \ref{ExperimentResults}.


\section{Experimental Study}
\label{ExperimentStudy}
In this section, we will introduce the processes of data collection and data preprocessing, and how to determine the optimum topic number in DTM.  
\subsection{Data Collection and Preprocessing}

The data collection process is described as follows. 
Firstly, we obtain tweet IDs from the  public available coronavirus Twitter dataset\footnote{https://github.com/echen102/covid-19-tweetids} that collected by Chen et al. \cite{chen2020covid} from Twitter API\footnote{https://developer.twitter.com/en/docs/api-reference-index} based on keywords such as Coronavirus, Covid, Covid19, Wuhanlockdown and account names such as CDCemergency, CDCgov, WHO, HHSGov to actively tracking tweets from Twitter. 
We collect totally 13,746,822 tweets with Tweepy (a python library for the Twitter API) based on the given tweet IDs  from April 1 to April 14.
After that, the non-English tweets in the tweet dataset are removed, and we get a total of 8,430,793 English tweets.

\begin{figure}[h]
    \centering
    \includegraphics[width=0.7\textwidth]{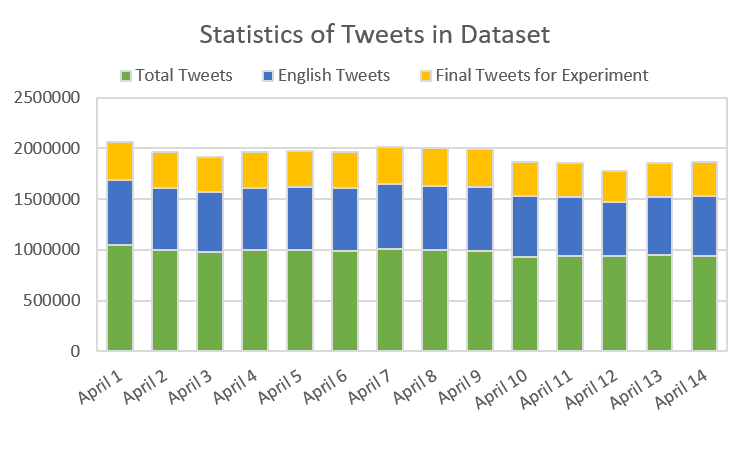}
    \caption{Statistics of tweets in dataset. The green rectangle represents the total tweets per day, the blue rectangle represents the English tweets in all tweets, and the yellow rectangle represents the pre-processed tweets used in the experiment.}
    \label{Statistics}
\end{figure}

Data preprocessing or cleaning is the first step for text mining task \cite{Yang2020}.
It includes converting all letters into lowercase, removing stop words, non-English letters, URLs, etc. 
Then, phrase extraction is used to ensure that words such as ``human\_rights" could be one token instead of separating ``human" and ``rights". Additionally, lemmatization is also adopted to remove inflectional endings and return the base or dictionary form of a word.
After preprocessing, we remove very short tweets (less than 6 words) and retained a total of 4,919,471 tweets with 269,391 unique tokens.
Figure \ref{Statistics} shows the ratio of raw tweets, English tweets, and finally adopted tweets. Table \ref{TweetNumberStastics} shows the number of daily tweets used for the experiment.
\begin{table}[h]
\centering
\caption{Size of daily tweets after preprocessing.}
\label{TweetNumberStastics}
\begin{tabular}{|l|c|c|c|c|c|c|c|}
\hline
Date  & April 1 & April 2 & April 3  & April 4  & April 5  & April 6  & April 7  \\ \hline
Total & 374,327 & 355,504 & 350,211  & 354,884  & 352,499  & 355,478  & 373,342  \\ \hline
Date  & April 8 & April 9 & April 10 & April 11 & April 12 & April 13 & April 14 \\ \hline
Total & 377,615 & 373,812 & 334,297  & 335,607  & 307,422  & 331,806  & 342,667  \\ \hline
\end{tabular}
\end{table}

\subsection{Topic Model Setup}
The number of topics is a crucial parameter in topic modeling and capable of making these topics human interpretable.
According to \cite{blei2006dynamic}, for the first slice of Dynamic Topic Models(DTM) to get setup, we fit LDA on the first day of the  dataset to learn the best topic number. 
We employ Gensim package in Python to train LDA model for the selection of  best topic number. 
Here, we use the coherence \cite{roder2015exploring}  by measuring the degree of semantic similarity between high scoring words of topics as an indicator to choose the best topic number. 
The coherence score helps distinguish between human understandable topics and artifacts of statistical inference.
\begin{align}
\label{coherence}
    Coherence=\sum_{i<j}score(w_i,w_j),
\end{align}

where select top $n$ frequently occurring words in each topic, then aggregate all the pairwise scores of the top $n$ words $w_i,\cdot\cdot\cdot,w_n$ of the topic. 

\begin{figure}[ht]
    \centering
    \includegraphics[width=0.7\textwidth]{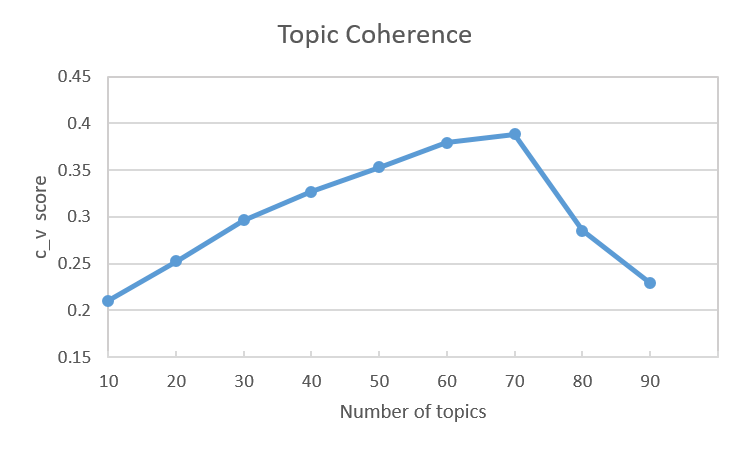}
    \caption{Coherence score of different topic numbers on LDA.}
    \label{Coherence}
\end{figure}

Figure \ref{Coherence} shows the coherence scores of different topic numbers on the LDA model based on one day tweets. 
As we can see, when the topic number is 70, the coherence score gets the maximum value that is around 0.39. Therefore, we assume the total number of topics is stable and set the topic number for DTM as 70.

\subsection{Results}
\label{ExperimentResults}

\begin{figure}[h]
    \centering
    \includegraphics[width=0.7\textwidth]{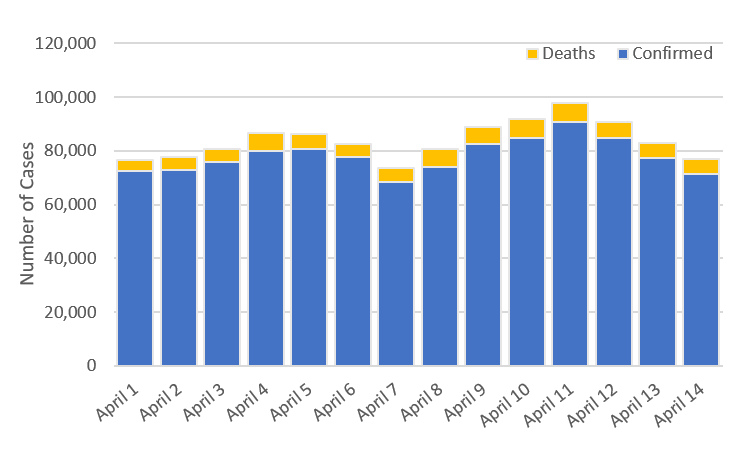}
    \caption{Worldwide deaths and confirmed cases of COVID-19 during the study period.}
    \label{WHOReport}
\end{figure}

Figure \ref{WHOReport} shows the overview of the number of deaths and confirmed cases, these datasets are collected from WHO \footnote{https://covid19.who.int/}. COVID-19 has gradually expanded to all parts of the world and has aroused the attention of countries, reported confirmed and deaths are slowly increasing.

\subsection{Overall Sentiment Dynamics}
\begin{figure}[!htbp]
\centering
\includegraphics[width=0.7\textwidth]{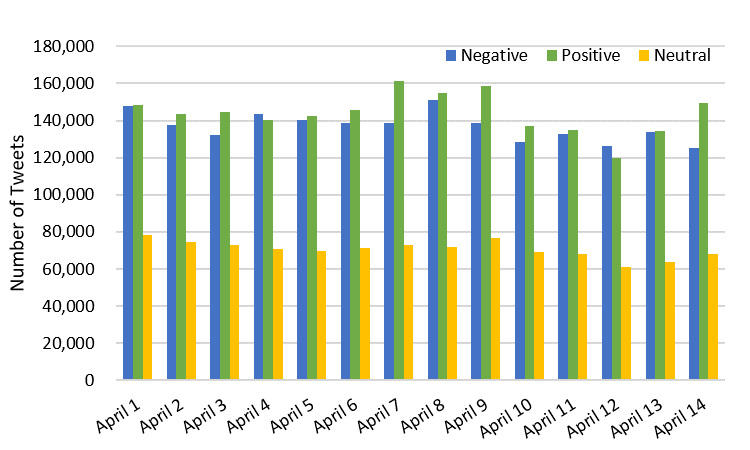}
\caption{The overall sentiment dynamics on Twitter during the study period.}
\label{DailyTweetsPolarityDistribution}
\end{figure}

Figure \ref{DailyTweetsPolarityDistribution} presents the overall sentiment distribution on Twitter during the study period. The number of tweets about the COVID-19 is around 350,000 per day, and the daily number of positive and negative tweets  is similar but all greater than the number of neutral tweets. 
In most days, the number of positive sentiment tweets is slightly higher than negative sentiment tweets. This shows that despite the spread of COVID-19, the community showed a dominant positive sentiment during the study period. This observation is also consistent with the finding of other researchers who have reported the similar conclusions based on country-level sentiment analysis \cite{bhat2020sentiment,jaidka2020estimating,zhou2020covid19}.

\subsection{Daily Hot Topics}
\begin{figure}[!htbp]
    \centering
    \includegraphics[width=115mm, scale=1.15]{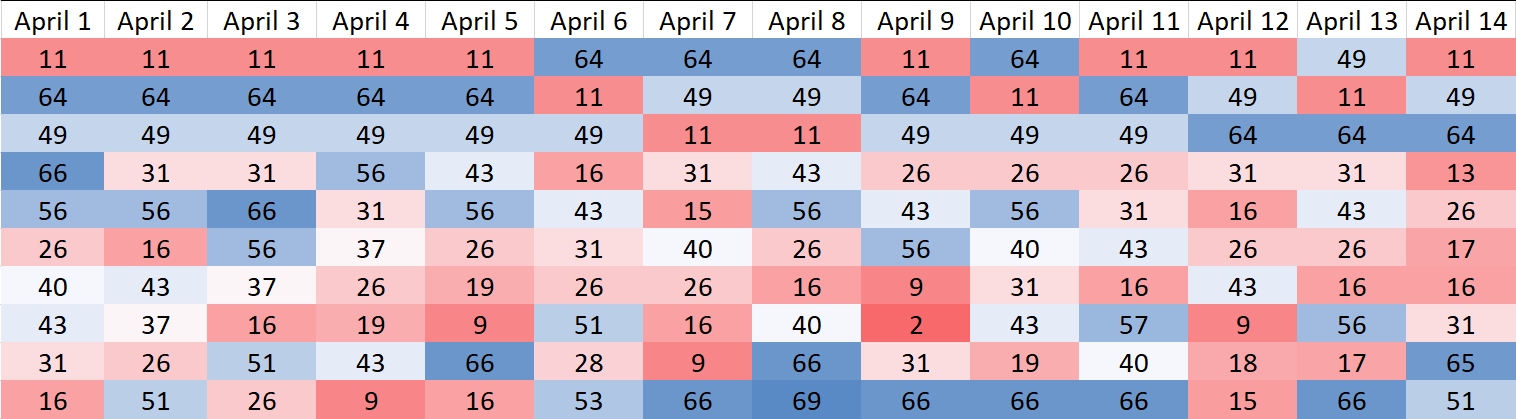}
    \caption{The top ten topics per day. \dahuihui{The number in the cell represents the index of the topic, and daily topics are sorted in descending order according to volume.}}
    \label{Index}
\end{figure}

\dahuihui{For daily topics discussed by users, we use DTM to generate 70 topics and then observe the popularity of topics}.
Table \ref{TopTen} shows the extracted top 10 highest volume topics with the most relevant words associated with the topics. To find the hottest topic discussed by people in the studied period, we rank the topics by their corresponding tweet number. 
Figure \ref{Index} shows the index of the top 10 topics for each day, topics are sorted by their volume from top row to bottom row in figure \ref{Index}. 
As we can see, topics 11, 49, and 64 are steadily ranked as the top 3 of daily hot topics.

\begin{table}[t]
\centering
\caption{Top 10 topics about COVID-19 on Twitter during the study period. \dahuihui{After each topic, the most contributing words related to the topic are displayed.} }
\label{TopTen}
\begin{tabular}{lllll}
\hline
Topic 64      & Topic 49            & Topic 11             & Topic 26          & Topic 31  \\ 
(disease)&(report)&(stay home)&(lockdown)&(life)\\
\hline
people       & case               & stay                & day              & time     \\ 
die          & new                & home                & lockdown         & good     \\ 
ignore       & death              & safe                & fight            & first    \\ 
seriously    & report             & go                  & covid            & life     \\ 
get          & total              & employee            & road             & talk     \\ 
disease      & important          & toxic\_relationship & go               & hard     \\ 
take         & far                & healthy             & medical\_supplie & hour     \\ 
go           & covid              & request             & deliver          & go       \\ 
say          & bank               & complete            & benefit          & failure  \\ 
intelligence & stuff              & panic               & stayhom          & pandemic \\ 
\hline
Topic 43      & Topic 16            & Topic 66             & Topic 56          & Topic 40 \\ 
(work)&(social\_distancing)&(stop spread)&(face mask)&(health care)\\
\hline
know         & think              & virus               & make             & health   \\ 
work         & thing              & spread              & mask             & care     \\ 
medium       & month              & stop                & face             & right    \\ 
create       & next               & woman               & share            & company  \\ 
lead         & social\_distancing & leader              & wear             & risk     \\ 
little       & go                 & move                & sell             & worker   \\ 
difficult    & vaccine            & citizen             & concern          & result   \\ 
street       & get                & act                 & stayhome         & would    \\ 
need         & article            & deadly              & expect           & demand   \\ 
tip          & finally            & slow                & wonder           & resource \\ \hline
\end{tabular}
\end{table}

\begin{figure}[!htbp]
    \centering
    \subfigure[Topic 11]{\label{wc_SATopic11}\includegraphics[width=0.33\textwidth]{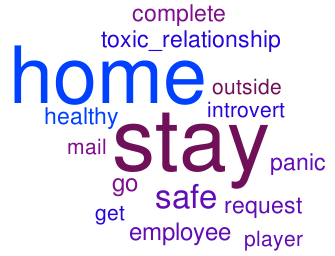}} 
    \subfigure[Topic 49]{\label{wc_SATopic49}\includegraphics[width=0.33\textwidth]{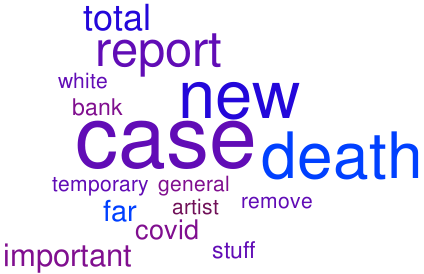}} 
    \subfigure[Topic 64]{\label{wc_SATopic64}\includegraphics[width=0.3\textwidth]{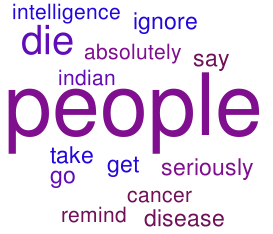}} 
    \caption{The most significant words for the three hot topics.}
    \label{wordcloud}
\end{figure}

Figure \ref{wordcloud} shows the most significant words that are associated with top three topics. Topics 11, 49, and 64 reflect the common concerns discussed by people, they are about staying at home to ensure safety, the latest case reports and \dahuihui{people death due to the disease}. 

To further analysis of these three topics, we hope to know people's opinions on these topics. 
We analyze the proportion of sentiment polarity of tweets under these topics to dynamically observe people's sentiment changes as the pandemic spreads, the results are shown in figures 9 to 11.
Overall, the topic's sentiment polarity various from topic to topic.

Topic 11 is related to staying at home, and our results show that the public kept a highly positive sentiment dominantly towards this measure to prevent infection. This may be because people work or study at home and enjoy more time with their families. They also had a positive belief to the combat against COVID-19 by the government and the society.

\begin{figure}[!]
    \centering
    \includegraphics[width=0.7\textwidth]{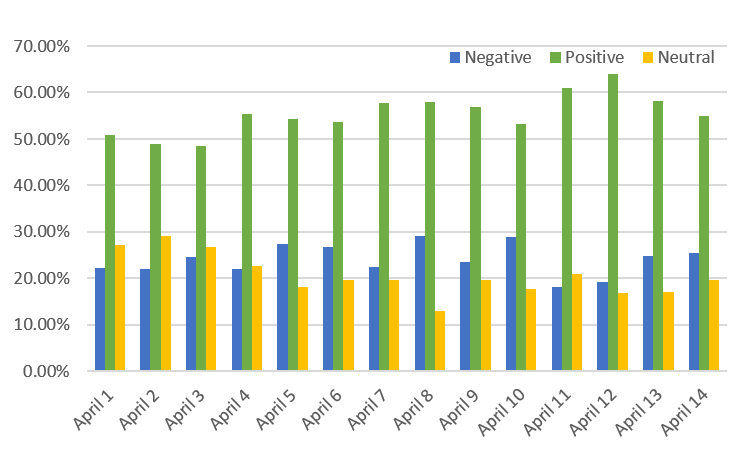}
    \caption{The sentiment dynamics of topic 11 during the study period.}
    \label{SATopic11}

\end{figure}

Topic 49 is about latest report about cases, the dominant sentiment around the topic of cases was almost negative despite positive sentiment existing. As the pandemic spread to more countries, the number of confirmed and deaths continues to rise, and people feel that the actual situation is worse than they expected. 
\begin{figure}[!]
    \centering
    \label{SATopic49}
    \includegraphics[width=0.7\textwidth]{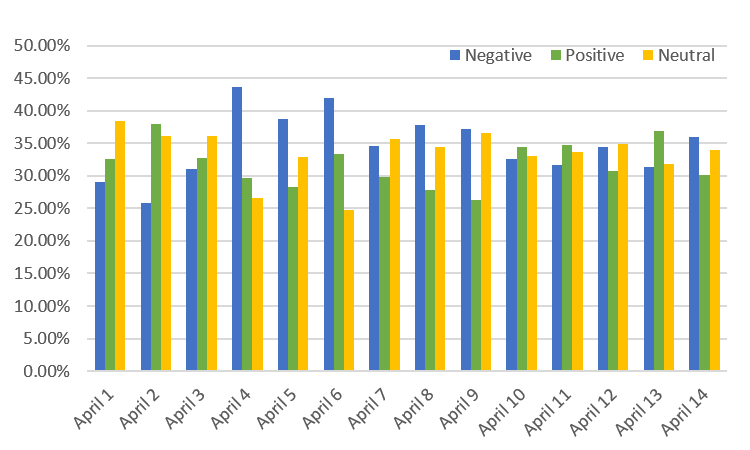}
    \caption{The sentiment dynamics of topic 49 during the study period.}
\end{figure}

Topic 64 \dahuihui{is about people's death due to COVID-19}, it shows a diametrically opposite result to topic 11, tweets with negative emotions are much higher than others. It is believed that the outbreak cannot be effectively controlled completely since policy makers ignored the seriousness of the pandemic, so they expressed strong dissatisfaction with this consequence. 

\begin{figure}[!]
    \centering
    \label{SATopic64}
    \includegraphics[width=0.7\textwidth]{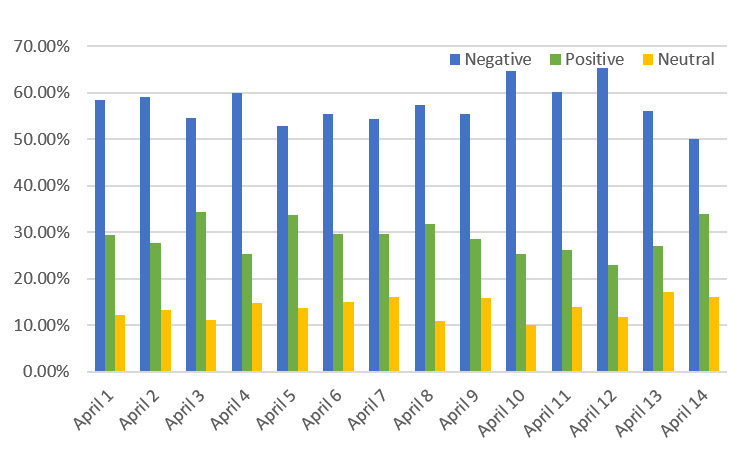}
    \caption{The sentiment dynamics of topic 64 during the study period.}
\end{figure}


\section{Conclusions and Future Work}
This study conducted a comprehensive analysis about hot topics with associated sentiment polarity distribution during COVID-19 period. Instead of the country-level study, the sentiment in this work was analysed at global-level on more than 13 million tweets collected from Twitter for two weeks from 1 April 2020. The results of analysis showed that people concern about the latest confirmed coronavirus cases, measures to prevent infection, the attitudes and specific measures of governments towards the pandemic.
The overall sentimental polarity was positive, but topic sentiment polarity various from topic to topic.  More interesting topics can be explored based on the current study in the future. For example, more specific topics can be analyzed to help policy maker, government and local communities to formulate measures to prevent the spread of negative emotions on social network, such as food shortage, lose job, debt, study at home.

\bibliographystyle{splncs04}
\bibliography{ADMA_BIB}

\end{document}